\documentclass[aps,prb,twocolumn,superscriptaddress,10pt]{revtex4-2}

\usepackage[dvipdfmx]{graphicx}
\usepackage{siunitx}
\usepackage{amsfonts}
\usepackage{amsmath}
\usepackage{amssymb}
\usepackage{bm}
\usepackage{graphicx}
\usepackage{dcolumn}
\usepackage{mciteplus}
\usepackage{euscript}
\usepackage{multirow}
\usepackage{color}
\usepackage{textcomp}
\usepackage{gensymb}
\usepackage{float}
\usepackage{enumitem}
\usepackage{xcolor}
\usepackage{sepfootnotes}
\usepackage{mhchem}
\usepackage{outlines}
\usepackage[hidelinks]{hyperref}

\graphicspath{{./}{./figs/}{../figs/}}

\DeclareSIUnit{\atomicunit}{a.u.}

\newcommand{\req}[1]{Eq.~\eqref{#1}}

\newcommand{\rfig}[1]{Fig.~\ref{#1}}
\newcommand{\rFig}[1]{Figure~\ref{#1}}
\newcommand{\rtbl}[1]{Table~\ref{#1}}

\newcommand{\rco}{R\text{Co}_5}
\newcommand{\rfe}{R\text{Fe}_{12}}
\newcommand{\rfeti}{R\text{Fe}_{11}\text{Ti}}
\newcommand{\rfeb}{R_2\text{Fe}_{14}\text{B}}

\newcommand{\eHartree}{E_{\rm H}}
\newcommand{\exc}{E_{\rm xc}}
\newcommand{\ylm}{Y_{lm}}
\newcommand{\ykq}{Y_{kq}}

\newcommand{\sieH}{\epsilon^{\rm H}}
\newcommand{\sieXC}{\epsilon^{\rm xc}}
\newcommand{\sieLDA}{\epsilon^{\rm LDA}}

\newcommand{\avg}[1]{\langle{#1}\rangle}

\newcommand{\mvm}[3]{\langle{#1}\lvert {#2} \rvert {#3}\rangle}
\newcommand{\st}[1]{\lvert{#1}\rangle}

\newcommand{\Oe}{\mathcal{O}}

\newcommand{\tbmnsn}{\text{TbMn}_6\text{Sn}_6}
\newcommand{\tbvsn}{\text{TbV}_6\text{Sn}_6}

\newcommand{\hCF}{H_\text{CF}}

\newcommand{\ud}{\mathrm{d}}

\newcommand{\udbra}{{\mathrm{d}}{\bf r}_1 \,}
\newcommand{\udbrb}{{\mathrm{d}}{\bf r}_2 \,}
\newcommand{\bfr}{{\bf r}}
\newcommand{\bfra}{{\bf r}_1}
\newcommand{\bfrb}{{\bf r}_2}

\begin{document}
\title{Toward a first-principles theory of rare-earth ions in crystals}

\author{Y. Lee}
\affiliation{Ames Laboratory, U.S.~Department of Energy, Ames, Iowa 50011}

\author{Z. Ning}
\affiliation{Ames Laboratory, U.S.~Department of Energy, Ames, Iowa 50011}

\author{R. Flint}
\affiliation{Ames Laboratory, U.S.~Department of Energy, Ames, Iowa 50011}
\affiliation{Department of Physics and Astronomy, Iowa State University, Ames, IA, 50011}

\author{R.~J.~McQueeney}
\affiliation{Ames Laboratory, U.S.~Department of Energy, Ames, Iowa 50011}
\affiliation{Department of Physics and Astronomy, Iowa State University, Ames, IA, 50011}

\author{I.~I.~Mazin}
\affiliation{Department of Physics and Astronomy, George Mason University, Fairfax, VA 22030}
\affiliation{Quantum Science and Engineering Center, George Mason University, Fairfax, VA 22030}

\author{Liqin Ke}
\email{liqinke@ameslab.gov}
\affiliation{Ames Laboratory, U.S.~Department of Energy, Ames, Iowa 50011}

\date{\today} 

\begin{abstract}
Density functional theory (DFT), including its extensions designed to treat strongly correlated localized electron systems such as DFT+$U$ and DFT+dynamical mean field theory, has proven exceedingly useful in studying the magnetic properties of solids.
However, materials with rare earths ($R$) have remained a notable exception.
The most vital rare-earth magnetic properties, such as magnetocrystalline anisotropy (MA), have been notoriously elusive due to the ubiquitous self-interaction error present in nearly all available DFT flavors.
In this work, we show explicitly how the orbital dependence of self-interaction error may contradict Hund's rules and plague MA calculations, and how analyzing DFT metastable states that respect Hund's rules can alleviate the problem.
We systematically investigate and discuss several rare-earth-containing families, $R$Co$_5$, $R_2$Fe$_{14}$B, $R$Fe$_{12}$, and $R$Mn$_6$Sn$_6$, to benchmark the MA calculations in DFT+$U$.
For all compounds we investigated, we found that our methodology reproduces the magnetic easy-axis, easy-plane, and non-trivial easy-cone anisotropies in full agreement with low-temperature experimental measurements.
Besides the fully-numerical ab initio approach, we further illustrate an efficient semi-analytical perturbation method that treats the crystal field as a perturbation in the limit of large spin-orbit coupling.
This approach evaluates the rare-earth anisotropy by assessing the dependence of crystal-field energy on spin-quantization axis rotation using $4f$ crystal-field levels obtained from non-spin-orbit calculations.
Our analytical method provides a quantitative microscopic understanding of the factors that control MA and can be used for predicting new high-MA materials.
\end{abstract}

\maketitle

\section{Introduction}

Among all the elements, the open-shelled lanthanides provide the largest magnetocrystalline anisotropy (MA), due to the strongly-localized nature of $4f$ orbitals and strong spin-orbit coupling (SOC), which can evolve substantially, including changing sign while varying the rare earth ($R$) element in an isostructural series of compounds.
The unparalleled strength and tunability of rare-earth MA allows for a wide range of applications, ranging from conventional high-performance permanent magnets~\cite{gschneidner1978book,schuler1979pb,lewis2013mmta,mccallum2014armr} to recent rare-earth-containing topological magnets~\cite{yin2020n,lee2023prb}.
To further exploit existing systems and explore new ones, \emph{ab initio} methods that can provide a microscopic understanding of rare-earth anisotropy and reliably predict new materials are highly desired.

The MA originates from the interplay between SOC and the crystal field (CF)~\cite{ke2015prb,ke2019prb}.
The $4f$ states are the \emph{most-localized} among all shells and generally well-shielded by the outermost electrons, resulting in a small CF splitting ($\Delta$) of tens of meV.
Considering the relatively large SOC strength $\xi$, CF effects can be treated as a perturbation, and the $4f$ orbital largely remains atomic-like.
The mechanism of $R$ MA can be understood in the following picture.
When the spin of $4f$ electrons rotates, in the first approximation, the charge of the strongly-correlated $4f$ electrons remains the same shape and follows the spin, as the spin and orbitals are locked by the large SOC.
The MA then arises from the energy variation corresponding to the rotating aspherical $4f$ cloud under the ligand-induced CF potential.
In the conventional CF theory, this energy dependence on spin direction $(\theta, \varphi)$ can be written as:
\begin{equation}
E(\theta,\varphi) = \int \ud \mathbf{r} n_{4f}(\mathbf{r}; \theta, \varphi) V_\text{CF}(\mathbf{r}) = \sum A_l^m Q_l^m(\theta,\phi) .
\end{equation}
Here, the CF potential of isostructural compounds is characterized by CF parameters (CFPs) $A_l^m$, while the asphericity of the rotating $4f$ charge, evolved with $4f$ orbital filling, is characterized by multipole moment $Q_l^m(\theta,\phi)$. The multipole moment can be expressed in terms of the Stevens coefficients $\Theta_l$, the operator equivalents $\Oe_l^m$, and the rare-earth radii $\avg{r^l}_{4f}$, e.g., $Q_l^0 = \Theta_l \avg{r^l}_{4f} \Oe_l^0$~\cite{hutchings1964ssp, taylor1972book, skomski1999book}.
Overall, the $4f$ electron configurations in solids, especially those of heavy $R$ elements, generally obey the same Hund's rules as in a free ion, according to the so-called standard rare-earth model (SRM)~\cite{jensen1991book, peters2014prb, locht2016prb}. The MA of $4f$ elements can reach the same order of magnitude as the CF strength, which typically ranges in tens of meV.

The atomic nature of the strongly-correlated, localized $4f$ electrons poses great challenges for mean-field methods such as density functional theory (DFT).
Various approaches, including the $4f$-open-core method, DFT+$U$~\cite{liechtenstein1995prb,lee2023prb}, dynamical mean-field theory (DMFT)~\cite{delange2017prb,pourovskii2020prb}, and quasiparticle self-consistent GW (QSGW)~\cite{chantis2007prb}, have been employed depending on the specific rare-earth properties being targeted.
DFT+$U$ is the simplest and most widely-used method to treat strong correlations.
Regarding the $4f$ MA, the primary issue with DFT+$U$ is that it is known to fail in reproducing the experimental ground-state $4f$ configuration~\cite{anisimov1993prb,czyzyk1994prb,solovyev1994prb,liechtenstein1995prb,shick1999prb,singh2006book,zhou2009prb}.
Specifically, it fails to reproduce Hund's second rule, which maximizes orbital polarization.

In general, DFT+$U$ can have many metastable $4f$-configuration solutions~\cite{gotsis2003prb,zhou2009prb,peters2014prb}, and the correct \textit{ground state} often appears in DFT+$U$ as a \textit{metastable} state that is hundreds of meV higher.
As discussed in detail, for instance, in Ref.~[\onlinecite{zhou2009prb}], the root of the problem is the orbital-dependent self-interaction error (SIE), stemming from the fact that each Kohn-Sham particle interacts with the \emph{total} charge density, including its own.
This orbital dependence of SIE is particularly significant for $4f$ orbitals, leading to incorrect orbital occupancies and $4f$ charge density, and consequently to incorrect MAE.

A key question arises: \emph{Can DFT+$U$ accurately describe the MA of tens of meV, even though it overestimates the energy of the true ground state by hundreds of meV?}
Our recent systematic study on topological magnetic compounds $R$Mn$_6$Sn$_6$ with heavy-$R$ elements has shown promise~\cite{lee2023prb}, provided that their Hund's rule ground states are enforced.
Not only are the easy directions of the entire series of compounds reproduced if Hund's rules are enforced~\cite{lee2023prb}, but the calculated MAE amplitude also agrees reasonably well with experiments~\cite{riberolles2022prx,rosenberg2022prb}.
However, it remains unclear how well the delicate MA in other rare-earth-containing magnets can be described using the SRM in the simplistic DFT+$U$ framework.
To better establish the validity and effectiveness of these methods, systematic investigations of MA in more rare-earth-based compounds are needed.

In this work, we first review and illustrate how the orbital dependence of SIE affects the $4f$ ground state and MA calculation in DFT-based methods.
We then discuss various methods that attempt to enforce Hund's rules, such as DFT+$U$, self-interaction corrections (SIC), and orbital polarization corrections (OPC), and how the additional terms therein affect the MA calculations.
We further systematically benchmark DFT+$U$ calculations of MA in several isostructural $R$-transition-metal ($R$-TM) intermetallic series, including well-established permanent magnet systems, $R$Co$_5$, $R_2$Fe$_{14}$B, and $R$Fe$_{12}$ with heavy $R$ elements.
In all cases, with the enforcement of Hund's rules, DFT+$U$ calculations provide a useful description of the MA without the need to include SIC and OPC corrections.
Finally, we demonstrate that the evolution of MA can be modeled purely analytically based on a perturbative treatment of the crystal field using the single-particle $4f$ levels obtained in DFT+$U$~\cite{lee2023prb}.

\section{SIE effects on 4f Ground state and MA in DFT: TbMn$_6$Sn$_6$ as an example}
Many-body effects are crucial for accurately describing the strongly-correlated $4f$ electrons.
Especially for light rare-earth elements, multiple Slater determinants are typically required to capture their complex electronic structure. 
Here, in this study, we focus primarily on the heavy $R$ elements with a large $R$-TM exchange coupling because their ground states effectively satisfy Hund's rules, and the $|L,S,J,m_J=J\rangle$ state with $J=L+S$ can, in principle, be  represented using a single Slater determinant, as in methods such as DFT~\footnote{It is worth noting that the \emph{exact} DFT is still a single-determinant method, yet it can describe MA. Theoretically, this is achieved through the Kohn-Sham wave functions that deviate from standard spherical harmonics in the single determinant. We will briefly discuss this issue later in the paper.}.
However, even for these ``relatively easier'' heavy-$R$ cases, challenges arise in describing $4f$ electrons, specifically related to the SIE and the corresponding violation of Hund's rules.

To gain a quantitative understanding of how SIE affects the ground state and MA, we illustrate this with a DFT+U calculation of $\tbmnsn$—a recently discovered quantum magnet with very strong easy-axis anisotropy.
According to Hund's rules, Tb$^{3+}$ ($4f^8$) is expected to have a fully-filled $4f$ majority-spin channel and one electron in the minority spin channel, with $4f^{1,\downarrow} _{\st{m_l = 3}}$.
This expectation is consistent with neutron scattering and magnetization measurements of $\tbmnsn$~\cite{riberolles2022prx} and $\tbvsn$~\cite{rosenberg2022prb}.
However, DFT+$U$ instead found a $4f$ ground state corresponding to $4f^{1,\downarrow} _{\st{m_l = 2}}$.
The experimental ground state is approximately $\Delta\epsilon = 340$ meV higher in $\tbmnsn$, appearing as a metastable state in DFT+$U$ (performed with SOC included and the experimental out-of-plane spin orientation at $U=\SI{10}{\eV}$).
Considering that the SOC included in calculation already lowers the $\st{3}$ state relative to $\st{2}$ by approximately $\frac{1}{2}\xi_{4f}^{\rm Tb} \approx$ \SI{120}{meV}, the orbital dependence of SIE for these two orbitals is about 460 meV, which is more than one order of magnitude larger than MA.

The SIE, while sizable, is practically independent of the crystallographic environment and is rotationally invariant.
The energy difference between these two $4f$ configurations remains essentially the same as for the free Tb$^{3+}$ ion, where we found $\Delta\epsilon_{\rm atom} = 350$ meV using a large supercell calculation.
Moreover, to ensure numerical accuracy, we calculated the variation of $\Delta\epsilon_{\rm atom}$ with spin rotation and found that the change is negligible.
In other words, the SIE is spin-rotationally invariant.

If, as we just established, the SIE is rotationally invariant, one may work around that by calculating the MA (and similar effects) not in the DFT ground state, but in a metastable state that respects Hund's rules.
This can be achieved by starting DFT+U calculations from a $4f$ occupation matrix constructed according to the desired orbital state, and by monitoring and controlling the orbital occupancy through the self-consistency process to ensure convergence closely to the targeted state.
Such capability is easy to implement and is generally available in popular DFT packages, including \textsc{Wien2k} and \textsc{Vasp}~\cite{Allen2014pccp}.

\begin{figure}[hbtp]
  \includegraphics[width=0.95\linewidth,clip,angle=0]{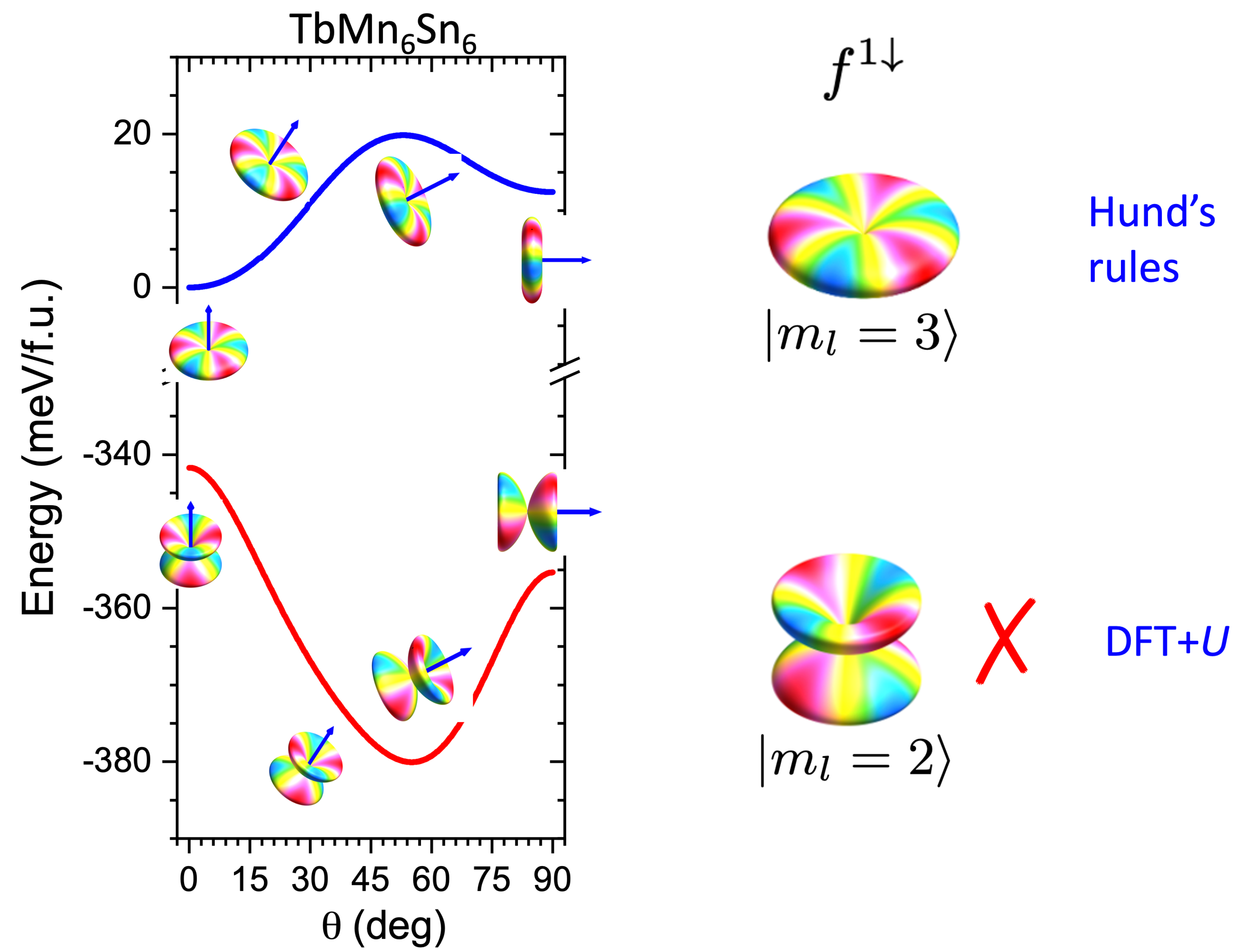}
  \caption{Magnetocrystalline anisotropy in TbMn$_6$Sn$_6$, represented by the variation of magnetic energy as a function of spin-axis rotation, calculated using DFT+$U$.
    The true ground state of Tb$^{3+}$ ($4f^8 = 4f^{7\uparrow} + 4f^{1\downarrow}$), following Hund's rules, appears as a metastable state in DFT+$U$.
    The anisotropy calculated for two configurations, the true ground state $4f^{1,\downarrow} _{\st{m_l = 3}}$ and the DFT+$U$ ground state $4f^{1,\downarrow} _{\st{m_l = 2}}$, is represented by the blue and red lines, respectively.
    The two $4f^{1,\downarrow}$ configurations are illustrated with polar plots of the corresponding complex spherical harmonics $Y_{l=3}^m(\theta,\phi)$, where the radius represents the amplitude and the color represents the phase at the point $(\theta, \phi)$.
  }
  \label{fig:figure1}
\end{figure}

\rFig{fig:figure1} shows the total energy variation as a function of the spin-quantization axis rotation, characterized by polar angle $\theta$, calculated for the two $4f$ configurations corresponding to the experimental and DFT ground states, respectively.
As illustrated in \rfig{fig:figure1}, at each polar angle, using the procedure discussed above, the calculations converge to solutions closely approximating the $4f^{1,\downarrow} _{\st{m_l = 3}}$ and $4f^{1,\downarrow} _{\st{m_l = 2}}$ configurations, respectively, in the local coordinate system (with the 
$z$-axis along the local spin direction).
The MA profiles calculated with these two solutions are markedly different.
Calculations using the $\st{m_l=2}$ DFT+U ground state yield an incorrect easy-cone MA, while those calculated with the $\st{m_l=3}$ configuration, the true ground state but metastable in DFT+U calculations, correctly host a strong easy-axis MA.

This is not surprising, as the $\st{m_l=2}$ and $\st{m_l=3}$ configurations lead to different asphericities of the $4f$ charge distribution, or equivalently, different multipole moments $Q_l$, which result in drastically different MA.
Therefore, for accurate MA calculations, it is crucial to enforce solutions that represent the correct $4f$ orbital configurations.

\section{Origin of erroneous Tb-$4f$ ground state: orbital dependence of SIE}

The origin of the erroneous $f^{1,\downarrow}_{\st{m_l = 2}}$ ground state in DFT calculations for the Tb$^{3+}$ ion is due to the strong orbital dependence of the SIE for $4f$ orbitals.
The Tb$^{3+}$ atom, with a $4f^8$ configuration, has a fully-occupied $4f$ majority-spin channel that produces an $s$-type spherical charge and potential.
In a single-particle Hamiltonian, without considering SOC, the seven $4f$ states should be degenerate if the potential is orbital-independent, as in plain DFT, and spherical.
Therefore, excluding self-interaction, the additional electron in the minority-spin channel, $f^{1,\downarrow}$, experiences a nearly spherical potential that does not lift the degeneracy of the seven $4f$ orbital states.
This is the same reason behind the well-known issue of $4f$ states being pinned at the Fermi level in DFT calculations unless a sizable Hubbard $U$ interaction is introduced in schemes such as DFT+U to polarize the occupied and unoccupied $4f$ states.
However, in DFT, the occupied $f^{1,\downarrow}$ electron generates an aspherical charge density that acts upon itself, as the functionals are evaluated using the total electron density.
The total SIE in the local density approximation (LDA), $\sieLDA$, originates from the Hartree energy, $\eHartree$, and the exchange-correlation energy, $\exc$, and can be written as
\begin{equation}
\sieLDA = \sieH + \sieXC,
\label{eq
}
\end{equation}
where $\sieH$ and $\sieXC$ are the corresponding SIE contributions associated with $\eHartree$ and $\exc$, respectively.
Due to the local approximation of the unknown exact exchange-correlation functional, $\sieH$ and $\sieXC$ do not cancel out as they do in the Hartree-Fock method, resulting in a nonzero $\sieLDA$.
Moreover, the orbital dependence of $\sieLDA$ is substantial for $4f$ states, leading to an incorrect $4f$ ground state.

Since the $4f$ charge asphericity and orbital dependence of SIE for the Tb$^{3+}$ ion (with $f^{7,\uparrow} + f^{1,\downarrow}$ configuration) are predominantly associated with the single electron in the minority-spin channel, we now present an analytical estimation of $\sieH$ and $\sieXC$ for the $f^1$ configurations with various $\st{m_l}$ states.
Obviously, we have $\sieH=\eHartree$ and $\sieXC=\exc$ for this single-electron model.
Here, we consider the eigenstates of the $f$ electron, where the angular part of the wavefunction is described by complex spherical harmonics $Y_{l=3}^m$.
As we will show, $\sieH$ favors the $\st{m_l=2}$ state, with the energy order $\st{2} < \st{1} < \st{3} < \st{0}$.
Conversely, $\sieXC$ favors the $\st{m_l=0}$ state, with the energy hierarchy $\st{2} > \st{1} > \st{3} > \st{0}$.
However, these contributions do not cancel each other out, resulting in an overall $\sieLDA$ that disfavors the $\st{m_l=3}$ state.

\subsection{Hartree self-interaction for $f^1$}
For the $f^1$ single-electron state, the $\sieH$ of the $\st{\pm m}$ state can be written as:
\begin{equation}
\label{eq:sieh}
\sieH_m = \frac{1}{2} \iint \udbra\udbrb \frac{\rho_m(\bfra) \rho_m(\bfrb)}{|\bfra - \bfrb|},
\end{equation}
where the electron density can be expressed in terms of the radial and angular parts of the wavefunction as $\rho_m(\bfr) = R_{4f}^2(r) |Y_{3m}(\theta, \phi)|^2$, with $m \in {0, 1, 2, 3}$.
The Coulomb interaction can be expanded using complex spherical harmonics as:
\begin{equation}
  \frac{1}{|\bfra-\bfrb|}=\sum_{k=0}^\infty\frac{r_<^k}{r_>^{k+1}} \frac{4\pi}{2k+1} \sum_{q=-k}^{q=k} \ykq(\theta_1,\phi_1) \ykq^*(\theta_2,\phi_2),
  \label{eq:coulomb_expand}
\end{equation}
where $\bfr_i = r_i (\sin\theta_i \cos\phi_i, \sin\theta_i \sin\phi_i, \cos\theta_i)$, and $r_< = \min(r_1, r_2)$ and $r_> = \max(r_1, r_2)$.

Substituting \req{eq:coulomb_expand} into \req{eq:sieh} and separating the radial and angular parts of the integration, we obtain:
\begin{equation}
\begin{pmatrix} 
\epsilon_0 \\ 
\epsilon_1 \\
\epsilon_2 \\
\epsilon_3 \\
\end{pmatrix}^{\rm H}
=F_0
+
\begin{pmatrix} 
 a_{02} & a_{04} & a_{06} \\ 
 a_{12} & a_{14} & a_{16} \\
 a_{22} & a_{24} & a_{26} \\
 a_{32} & a_{34} & a_{36} \\ 
\end{pmatrix}
\begin{pmatrix} 
F_2 \\
F_4 \\
F_6 \\
\end{pmatrix}
\approx
F_0 +
\begin{pmatrix} 
 \tilde{a}_{02} \\ 
 \tilde{a}_{12} \\
 \tilde{a}_{22} \\
 \tilde{a}_{32} \\ 
\end{pmatrix}
F_2.
\label{eq:sieh_3}
\end{equation}

Here, the radial integrals are represented by the Slater integrals $[F_0, F_2, F_4, F_6]$, and the angular integrals are represented by the matrix element $a_{mk}$, which can be evaluated using the Gaunt coefficients as follows:
\begin{equation}
a_{mk} = \frac{4\pi}{2k+1} \left[ \operatorname{Gaunt}(3, k, 3; -m, 0, m) \right]^2.
\label{eq:amk}
\end{equation}
The last step of \req{eq:sieh_3} is obtained using $a_{m0}=1$ and assuming $F_4/F_2\approx0.6681$ and $F_6/F_2\approx0.4943$.

\begin{table}[ht]
  \caption{ The coefficients $a_{mk}$, where $m$ denotes the complex spherical harmonics.
The coefficients $\tilde{a}_{m2}=a_{m2}+0.6681 a_{m4}+  0.4943a_{m6}$ is calculated by assuming the ratio between $F_2$, $F_4$, and $F_6$ values as $1:0.6681:0.4943$.
The orbital-dependent part of $\sieH_m$, $\Delta\sieH_m$ (in units of meV), is then calculated by further assuming $F_2=10$~eV, with the value of the $\st{\pm2}$ state as the reference zero.
For complex $\st{m}$ states, the ordering is $\st{\pm2}<\st{\pm1}<\st{\pm3}<\st{0}$.
}
\label{tbl:amk}%
\bgroup
\def\arraystretch{1.2}
\begin{tabular*}{\linewidth}{c @{\extracolsep{\fill}} cccccc}
\hline\hline
     $m$      &   &  $a_{m2}$  &  $a_{m4}$ &  $a_{m6}$ &  $\tilde{a}_{m2}$ & $\Delta\sieH_m$     \\ \hline
      0       &   &  0.0711    &  0.0331   &  0.0543   &   0.1201          & 876     \\ 
      $\pm1$  &   &  0.0400    &  0.0009   &  0.0306   &   0.0557          & 232     \\ 
      $\pm2$  &   &  0         &  0.0450   &  0.0049   &   0.0325          & 0       \\ 
      $\pm3$  &   &  0.1111    &  0.0083   &  0.0001   &   0.1167          & 842     \\ 
    \hline\hline
\end{tabular*}
\egroup
\end{table}

\rtbl{tbl:amk} lists the matrix elements $a_{mk}$ and the effective element $\tilde{a}_{m2}$, as well as the orbital-dependent part of $\sieH_m$, $\Delta\sieH_m\approx\tilde{a}_{m2}F_2$, calculated with $F_2=10$~eV, with respect to the $\st{m_l=2}$ state.
Clearly, $\sieH$ favors $\st{m_l=\pm2}$ states while disfavoring $\st{m_l=0}$ and $\st{m_l=\pm3}$ states.
The small SIE of $\st{\pm2}$ is due to the vanishing of the matrix element $a_{22}$ calculated using \req{eq:amk}, which results from the fact that they satisfy one of the conditions for non-trivial zeros of Wigner-$3j$ symbols, i.e., $\operatorname{Wigner3j}(3,k,3;m,0,-m)=0$ with $k=2$.

\subsection{Exchange-correlation self-interaction for $f^1$}
For the LDA exchange-correlation energy $\exc[\rho] \propto -\int \rho^{\frac43}(\bfr) \ud r$, the orbital-dependent $\sieXC_m$ for the $f^1=\st{\pm m}$ states is characterized by the angular part of the integration, 
\begin{equation}
\Omega(\sieXC_m) = -\int\ud\Omega \left(|\ylm(\theta,\phi)|^2\right)^{\frac43}.
\end{equation}


\begin{table}[ht]
  \caption{Angular part integration of $\sieXC$, denoted as $\Omega(\sieXC_m)$, for the $f^1$ configuration.
$\Delta\Omega(\sieXC_m)$ represents the $\Omega(\sieXC_m)$ values with respect to the $\st{m=0}$ state.
}
\label{tbl:elda}%
\bgroup
\def\arraystretch{1.2}
\begin{tabular*}{\linewidth}{c @{\extracolsep{\fill}} rrrr}
\hline\hline
    $m$                  & 0        & $\pm1$  & $\pm2$  & $\pm3$   \\ \hline
$\Omega(\sieXC_m)$       & -0.5314  & -0.4903 & -0.4801 & -0.4963  \\
$\Delta\Omega(\sieXC_m)$ & 0        &  0.0411 &  0.0513 &  0.0351  \\
    \hline\hline
\end{tabular*}
\egroup
\end{table}
\rtbl{tbl:elda} lists $\Omega(\sieXC_m)$ values and the corresponding values with respect to the $\st{m=0}$ state.
Clearly, $\sieXC_m$ favors the $\st{0}$ state, with the energy order $\st{0}<\st{\pm3}<\st{\pm1}<\st{\pm2}$.
Numerically, our DFT+$U$ calculations for the Tb$^{3+}$ free ion also shows that $\st{2}$ has the highest $\sieXC$, while $\st{0}$ has the lowest $\sieXC$, consistent with this finding.

Overall, when combining $\sieH$ and $\sieXC$, the total $\sieLDA$ yields a much higher energy for $\st{\pm3}$ solutions compared to other $\st{m}$ solutions.
Specifically, $\sieH$ strongly favors $\st{\pm2}$ much more than $\st{0}$ and $\st{\pm3}$, while $\sieXC$ favors $\st{0}$.
Consequently, $\sieLDA$ results in a significantly higher energy for $\st{\pm3}$ states than for other states.
The SOC energy, on the other hand, favors states with large positive $m_l$ values in the minority-spin channel.
While it may not be sufficient to overcome the SIE to stabilize the true ground state of $\st{3}$, it does lower the energy of the $\st{2}$ state below that of the $\st{0}$ and $\st{1}$ states, resulting in an erroneous ground state of $f^{1,\downarrow} _{\st{m_l = 2}}$ in calculations.

\section{DFT+$U$, SIC, and OPC}
Various methods have been developed and employed to improve the DFT description of $4f$ electrons, including DFT+U, SIC, and OPC methods.
Both SIC~\cite{janesko2022jcp} and OPC~\cite{solovyev1998prl} methods can be connected to the more general DFT+U method.
In this section, we discuss their applications to MA calculations.

\paragraph*{DFT+U approach.} To resolve the unphysical pinning of $4f$ states near the Fermi level in DFT, DFT+$U$ with a sizable Hubbard $U$ value is the most employed method to treat the well-localized $4f$ orbitals, shifting the occupied and unoccupied $4f$ states away from the Fermi level by $\pm \frac12(U-J_\text{H})$, respectively.

The DFT+$U$ total energy, which differs from the plain DFT one by a correlation contribution from the Hubbard-type model Hamiltonian for the selected orbitals, can be written as
\begin{equation}
E_{{\rm LDA}+U}[\rho(\bfr),\mathbf{n}] = E_{\rm LDA}(\rho) + E^{\rm corr}(\mathbf{n}),
\label{eq
}
\end{equation}
where the correlation energy is evaluated using the occupation matrix $\mathbf{n}$ with the screened Coulomb interactions parameterized with U and J values as
\begin{equation}
E^{\rm corr}(\mathbf{n}) = E^{\rm Hub}(\mathbf{n}) - E_{\rm dc}(n).
\label{eq
}
\end{equation}
Here, the Hartree-Fock-like interaction $E^{\rm Hub}(\mathbf{n})$ is self-interaction-free as the SIE of the direct and exchange terms is exactly canceled out~\cite{shick1999prb}; the double-counting term $E_{\rm dc}(n)$, which accounts for the interaction already included in LDA, is not uniquely defined and depends on the implementation scheme.
Typically, it depends only on the trace of $\mathbf{n}$; therefore, $E_{\rm dc}(n)$ depends only on the number, but not the orbital character $m_l$, of the occupied states.
It is worth noting that, besides of the aforementioned splitting between the occupied and unoccupied $4f$ states by $(U-J_\text{H})$, in the popular fully-localized-limit (FLL) double-counting scheme, $J_{\rm H}$ also induces the spin splitting of corresponding $4f$ levels, depending on the orbital's occupancy. 
Overall, the $E^{\rm corr}(\mathbf{n})$ in DFT+U, which consists of the SIE-free $E^{\rm Hub}(\mathbf{n})$ and the orbital-independent $E_{\rm dc}(n)$, do not explicitly address the orbital dependence of SIE.
Therefore, the SIE inherited from the original DFT in DFT+$U$ remains largely intact, and conventional DFT+U schemes are not expected to correct the Tb$^{3+}$ ground state discussed above.

Alternative DFT+U schemes that aim to minimize the orbital dependence of SIE have been proposed.
An interesting work by Zhou and Ozoliņš modifies only the exchange term of the LDA by including only the exchange, but not Hartree, component of $E^{\rm corr}(\mathbf{n})$.
The exchange-only $E^{\rm corr}(\mathbf{n})$ now contains orbital-dependent SIE and can be used to minimize the orbital dependence of SIE by properly mixing the FLL $E_{\rm dc}(n)$ exchange and LDA exchange.
This method has been demonstrated to improve the description of the $4f$ ground-state and other properties such as CFP and optical properties~\cite{zhou2009prb,zhou2012prb}.
However, such corrections, with rotational invariant $E^{\rm corr}(\mathbf{n})$, does not explicitly affect the calculated $E(\theta,\phi)$ profile once the $4f$ configuration is enforced during the rotation.

Therefore, the main effect of applying the $U$ parameter is to shift the occupied 4f states away from the Fermi level.
This shift is necessary to be consistent with experiments and helps ensure convergence to the desired 4f orbital occupation that respects all three Hund's rules, which may otherwise be disrupted by strong hybridization between 4f and ligand orbitals.
If a sizable $U$ is chosen and the hybridization between 4f and ligand orbitals is negligible, the electron correlation induced by DFT+U does not explicitly affect the anisotropy calculation.
This is because, when the spin-quantization axis rotates, the $U$- and $J_{\rm H}$-dependent correlation energy remains constant as long as the orbital occupancy remains the same in the local coordinate system.
On the other hand, in the range of $U$ values that lead to strong hybridization between 4f and ligand orbitals, a much stronger $U$ dependence of MA is expected.
This is because the contribution of hybridization, in addition to the crystal electric field, becomes more significant for MA.

\paragraph*{Self-interaction correction.}

The SIC method, proposed by Perdew and Zunger in 1981~\cite{perdew1981prb}, was initially inspired by the problem of reproducing the correct energy gap in insulators.
They pointed out that in the limit of one-electron systems, the exchange-correlation potential should exactly cancel the Hartree potential, which was not the case for all functionals available at that time.
It was further believed that the fundamental gap is a ground-state property and thus must be reproduced in exact DFT.
With this in mind, Perdew and Zunger proposed a method that deducts the self-interaction energy of each orbital from the DFT functional.
The resulting orbital-dependent functional was neither a Kohn-Sham functional nor uniquely defined for many-electron systems~\cite{perdew1981prb,svane1988prb,zhang1998jcp}.
Nevertheless, it was conceived that this functional would offer a better approximation to the exact Kohn-Sham functional than existing local flavors.
While the functional was shown to considerably improve the excitation gap, it was never proven to systematically improve the total energy.
Two years later, it would be proven~\cite{perdew1983prl,sham1983prl} that the fundamental gap is not a ground-state property and need not (and is unlikely) to be reproduced in exact DFT.
It was argued~\cite{maksimov1989jpcm} that, in reality, this method is not necessarily a superior approximation to Kohn-Sham DFT, but rather a fortuitously good approximation to the Dyson equation for one-particle excitations.
Indeed, the weighted density functional~\cite{alonso1978prb}, which is inherently self-interaction-free in the Perdew-Zunger sense and yields improved total energy and linear response~\cite{mazin1998prb}, produces results that are quite different from those of SIC LDA or GGA functionals.
Thus, there is no solid foundation for expecting that such non-DFT SIC functionals would universally account for the Hund's rules in f-electron systems, nor is this method (as opposed to DFT+U) commonly implemented in modern DFT codes.

\paragraph*{Orbital polarization correction.}

In analogy to the Stoner expression for spin polarization $-\frac{1}{4}IM_s^2$, Brooks and coworkers~\cite{eriksson1990prb} introduced an orbital polarization term proportional to $-\frac{1}{2}L^2$, giving rise to a one-electron eigenvalue shift ($-E^3Lm_l$) for the state $\st{m_l}$.
Here, the Racah parameter $E^3$, which can be related to Slater integrals ($F_2$, $F_4$, and $F_6$), plays a role analogous to the Stoner $I$ for spin polarization.

While this method does introduce a correction that tries to maximize the orbital moment and, thus, technically can enforce Hund's rules, it has no direct first-principles justification.
Various attempts~\cite{eschrig2005el,solovyev1998prl} to derive {\it an} OPC have resulted in formulations that, while potentially useful, differ from the suggested form.
To the best of our knowledge, the more elaborate OPC schemes beyond the original description of Brooks and coworkers are neither implemented in standard codes nor universally tested.

The original OPC prescription is implemented in \textsc{Wien2k} code and we applied it to $\tbvsn$.
It appears that achieving the Hund’s rule state using the OP method is quite challenging, if not impossible.
In the minority spin channel, $\st{m_l=2}$ levels remain the lowest unless a very high OP parameter is applied to promote the occupation of the $\st{m_l=3}$ state.
However, since the orbital polarization term is spin-independent, such a large OP parameter also causes large orbital polarization in the majority-spin channel, resulting in partial occupation in the majority-spin channel.
As a result, with this OPC scheme, we are not able to obtain the correct $4f$ ground state that satisfies Hund's rules.
Thus, we conclude that neither the SIC nor the OPC method, at least by itself, is useful for extensive calculations of MA in $4f$-metal compounds.
Therefore, we will pursue the idea discussed above of calculating MA in an artificially stabilized, computationally-metastable orbital state that respects Hund's rules.

\section{$4f$ anisotropy: Benchmarking Total Energy Calculation}
To systematically benchmark the validity of MA calculations, we further investigate several isostructural systems, including the two most important permanent magnet systems: $\rco$ and $\rfeb$.
Among them, SmCo$_5$- and Fe-rich Nd$_2$Fe$_{14}$B-based magnets are the most successful permanent magnets so far.
We will show that the rare-earth MA in these systems can be well described using DFT+$U$.

Various methods, including DFT+$U$, SIC~\cite{patrick2018prb}, and DMFT in the form of Hubbard I~\cite{delange2017prb,pourovskii2020prb}, have been employed to investigate the rare-earth MA in these systems, especially for SmCo$_5$ due to its importance and a smaller $\rco$ unit cell.
However, despite the wide application of simplistic DFT+$U$, the systematic MA study of isostructural series with heavy-$R$ elements is, to the best of our knowledge, rare.
Moreover, most of the previous calculations in the literature did not discuss the details of the converged $4f$ configuration or were carried out without enforcing Hund's rules; the calculated orbital moments can deviate significantly from SRM due to the orbital-dependent SIE in DFT+$U$ and the corresponding failure to reproduce Hund's rules being ignored.
Consequently, the reported orbital magnetic moment and MA values are scattered and hard to evaluate, casting doubt on the validity of DFT+$U$ applications for rare-earth MA.

Therefore, here we want to fill this gap by systematically benchmarking MA calculations with the SRM model using DFT+$U$.
Such benchmarking is also necessary if we want to compare with more sophisticated approaches such as DMFT or other methods and evaluate their improvement.

Here, we focus on the rare-earth MA in these systems, although the transition-metal sublattice MA is also important and of interest by itself~\cite{daalderop1996prb}.
For example, in $\rco$, the Co sublattice also contributes a large easy-axis anisotropy, as YCo$_5$ represents one of the largest $3d$ MA systems.
However, plain DFT underestimates the MA of Co sublattices and only gives a value between $\frac{1}{4}$ and $\frac{1}{3}$ of the experimental value in $\rco$~\cite{ke2016prb}.
Orbital polarization~\cite{steinbeck2001jmmm,soderlind2017prb} or applying an additional Hubbard $U$ interaction on Co-$3d$ orbitals in DMFT~\cite{zhu2014prx} or DFT+$U$ has been used to improve the agreement between calculation and experiments.

\paragraph*{Computational details.}
DFT+$U$ calculations were conducted using \textsc{Wien2K}.
The only constraint we enforced was Hund's rules at the local coordinate.
We rotated the spin axis while maintaining the correct occupied orbital states and calculated the MA as the variation of energy with respect to spin axis rotation.
A sizable $U$ is necessary to ensure that the $4f$ states can converge to the Hund's rules state; otherwise, hybridization may not allow for it.
The calculated MA is not very sensitive to $U_f$, as long as it is large enough to ensure that the $f$-states are well-removed from the Fermi energy, and the orbital configuration of the occupied $f$-state respects all of Hund's rules as in SRM.
Increasing the $U$ parameter further usually has a smaller effect on energy because the hybridization is already small, and the charge densities of $4f$ and ligands do not change significantly with $U$.
Additionally, we performed MA calculations for corresponding Gd compounds (or treated $4f$ as a spherical open-core) to obtain the non-$4f$ contributions to the total MA~\cite{lee2023prb}.

\subsection{$\rco$}

\begin{figure}[htb]
\centering
\begin{tabular}{c}
  \includegraphics[width=.70\linewidth,clip]{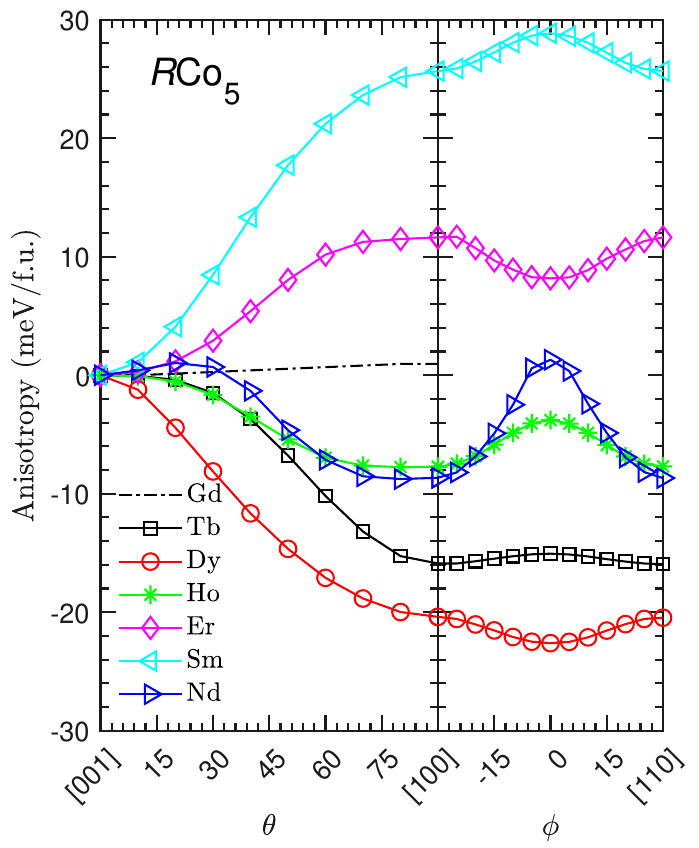}    \\
\end{tabular}%
\caption{
Variation of magnetic energy (in meV/f.u.) calculated in DFT+$U$ as a function of spin-axis rotation in $\rco$ with $R =$ Tb, Dy, Ho, Er, Sm, and Nd. The spin direction is characterized by the polar angle $\theta$ and the azimuthal angle $\phi$. The lattice vector $\mathbf{c}$ ([0 0 1]) direction is along the $\hat{z}$ direction and is denoted by $\theta=0\degree$, while the lattice vector $\mathbf{a}$ ([1 0 0]) direction is denoted by $\theta=90\degree$ and $\phi=-30\degree$. The calculations are performed in DFT+$U$ with $U\approx0.7$ Ry on the $4f$ states of all $R$ elements to satisfy Hund's rules. For all the depicted compounds, the calculated easy directions are consistent with experimental observations.
}
\label{fig:total_e_ma_15}
\end{figure}

\rFig{fig:total_e_ma_15} shows the calculated total energies $E(\theta)$ in $\rco$ as functions of spin-quantization direction characterized by the polar angle $\theta$ and the azimuthal angle $\phi$.
Besides the heavy $R$ elements, we also consider $R=$ Sm and Nd for comparison with existing experimental data.
In contrast to other $\rco$ compounds, GdCo$_5$ with a spherical Gd-$4f$ charge exhibits a very small easy-axis MA, contributed mostly by the Co sublattices.
The energy minimum occurs at $\theta=0\degree$, [0 0 1], for Er and Sm, and at $\theta=90\degree$ for all other compounds.
This suggests that $\rco$ has an easy-axis MA for $R=$ Er and Sm, while an easy-plane MA for $R=$ Tb, Dy, Ho, and Nd.
The calculations for all the compounds accurately reproduce their experimental easy directions measured at low temperatures~\cite{kelaeev1983pss, zhao1991prb, skokov2011jmmm, larson2004prb}, demonstrating the effectiveness of MA description in SRM through DFT+$U$.

HoCo$_5$ shares a similar MA profile with NdCo$_5$ but has an opposite MA profile to ErCo$_5$.
This can be understood as Ho$^{3+}$ with a $4f^{3,\downarrow}$ configuration and Nd$^{3+}$ with a $4f^{3,\uparrow}$ configuration having a similar aspherical charge density in the single-Slater-determinant description of DFT, if one ignores the difference between their radial wavefunctions.
The nearly perfect opposite MA profiles of HoCo$_5$ and ErCo$_5$ reflect the particle-hole symmetry also found in HoMn$_6$Sn$_6$ and ErMn$_6$Sn$_6$~\cite{lee2023prb}.

Interestingly, all $\rco$ compounds exhibit a sizable in-plane MA, suggesting a significant higher-order CFP $A_6^6$.
Among all $R$ elements, TbCo$_5$ has the smallest in-plane MA, while NdCo$_5$ shows the strongest in-plane MA, almost equal in amplitude to the out-of-plane MA.
The large in-plane MA in NdCo$_5$ is consistent with previous experiments~\cite{ermolenko1976ieee, ermolenko1980pssa} and has also been reproduced in a recent DMFT study~\cite{pourovskii2020prb}.
Assuming a fixed CFP $A_6^6$ for the isostructural $\rco$, the magnitude of in-plane MA correlates well with the element's multipole moment $Q_6$, with the largest value found in Nd and the smallest in Tb.
It is worth noting that the in-plane MA in SmCo$_5$ would vanish in a conventional CFP model using the lowest multiplet $\st{L=5,S=\frac{5}{2},J=\frac{5}{2},m_J}$, as the Stevens operator $O_6^6$ vanishes for $J=\frac{5}{2}$, unless the $J$ mixing due to multiplet interaction is considered.
The non-zero in-plane MA reflects a difference between the many-body treatment and the single-Slater-determinant description of DFT for the Sm ion.

\subsection{$\rfeb$}

\begin{figure}[htb]
\centering
\begin{tabular}{c}
  \includegraphics[width=.70\linewidth,clip]{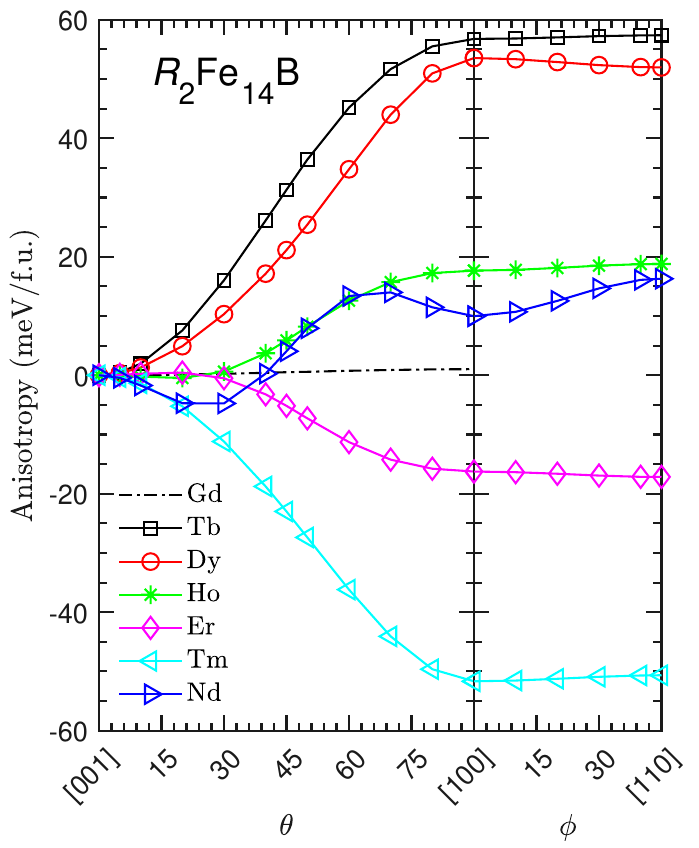} \\
\end{tabular}%
\caption{
  Variation of magnetic energy (in meV/f.u.) calculated in DFT+$U$ as a function of spin-axis rotation in $\rfeb$ with $R =$ Tb, Dy, Ho, Er, Tm, and Nd.
  The spin direction is characterized by the polar angle $\theta$ and the azimuthal angle $\phi$.
  The lattice vector $\mathbf{c}$ ([0 0 1]) direction is along the $\hat{z}$ direction and denoted by $\theta=0\degree$, while the lattice vector $\mathbf{a}$ ([1 0 0]) direction is denoted by $\theta=90\degree$ and $\phi=0\degree$. 
The calculations are performed in DFT+$U$ with $U\approx0.7$ Ry on the $4f$ states of all $R$ elements to satisfy Hund's rules.
For all the depicted compounds, the calculated easy directions are consistent with experimental observations.
}
\label{fig:total_e_ma_2141}
\end{figure}

$\rfeb$ compounds crystallize in a tetragonal crystal structure with space group $P4_2/mnm$ ($\#136$).
There are two inequivalent $R$ sites, denoted by Wyckoff sites $4g$ and $4f$.
The primitive cell consists of four formula units.
Experimentally, the easy directions of $\rfeb$ at low temperatures are easy-axial for Tb and Dy, conical for Ho and Nd, and easy-plane for Er and Tm~\cite{szytula1994book}.

\rFig{fig:total_e_ma_2141} shows the MA calculated in $\rfeb$ with the spin quantization direction rotating from $[001]$ to $[100]$ and then to $[110]$ directions of the tetragonal crystal structure.
The calculated easy directions again all agree with experimental observations.
However, for the in-plane MA, $\rfeb$ compounds show somewhat smaller values than those in $\rco$.

Experimentally, it was found that the net magnetization in Nd$_2$Fe$_{14}$B cants away from the $c$ axis toward the [110] direction by an angle of $\theta=30\degree$, measured at $\SI{4}{\kelvin}$.
This is consistent with the calculated energy minimum occurring at $\theta=30\degree$ when the spin rotates from [001] to [100], as shown in \rfig{fig:total_e_ma_2141}
We further confirm that rotation from [001] toward [110] produces a slightly deeper energy minimum at $\theta=30\degree$ (not shown), thus reproducing exactly the experimental easy-cone angle.
The contribution from the two inequivalent Nd sites to the MA is also of great interest.
It has been argued that the $4f$ and $4g$ sites have negative and positive contributions, respectively, to the MAE~\cite{haskel2005prl}.
However, we found that contributions from both sites show an energy minimum near $\theta=30\degree$. 

Remarkably, very strong easy-axis MA is obtained for $R$=Tb and Dy.
In fact, in practice, a small amount of these two heavy $R$ elements is often required to enhance the coercivities of $\rfeb$-based magnets for real applications.
Similar to $\rco$, the calculated $E(\theta)$ profiles of Ho$_2$Fe$_{14}$B and Er$_2$Fe$_{14}$B MA also show perfect particle-hole symmetry.

\subsection{Other isostructural series}

\paragraph*{$R$Fe$_{12}$ compounds.} Fe-rich $\rfe$-based compounds have recently attracted significant interest in the permanent magnet community~\cite{ke2016prbA}.
In general, these compounds typically form as $R$Fe$_{12-x}M_{x}$, requiring a third element $M$ = Ti, V, Cr, Mn, Mo, W, Al, or Si to stabilize the structure. Experimental easy-axis information for $R$Fe$_{11}$Ti is available for comparison, though there are some disagreements in experimental reports~\cite{boltich1989jmmm,kou1993prb,hu1989jpc,szytula1994book}. For example, both easy-plane and easy-cone MA have been reported for $R$ = Tb, while both easy-axis and easy-cone MA have been reported for $R$ = Ho at low temperatures.

To compare with experiments, we calculated the MA in the hypothetical composition of $\rfe$, ignoring the third element for simplicity. 
We found that the calculated MAE per $R$ atom of $\rfe$ is more than five times smaller than in $\rco$ and $\rfeb$.
The calculated MA is easy-plane for $R$ = Tb and Dy, and easy-cone for $R$ = Ho, which agrees with the experimental findings reported~\cite{boltich1989jmmm,szytula1994book}.
For $R$ = Er and Tm, however, our calculated easy directions for $\rfe$ do not exactly match the experimental results for $\rfeti$.
Experiments found that the MA is easy-cone and easy-axis, respectively, for $R$ = Er and Tm in $\rfeti$.
In contrast, our calculations show local minima for these experimental easy directions, but they are slightly higher (about 0.24 and 0.3 meV per $R$ atom, respectively) than the in-plane direction.
The discrepancy is likely due to the presence of the Ti atom in the real materials; the chemical effect and induced crystal structure distortion can modify the crystal field of the $R$ element and MA~\cite{patrick2024prl}.
More comprehensive experiments and MA calculations with more realistic crystal structures and compositions are still needed to elucidate MA in $\rfe$-based systems.

\paragraph*{$R$Mn$_6$Sn$_{12}$ and $R$V$_6$Sn$_{6}$ compounds.}
Besides these three permanent magnet systems, we have also previously investigated the rare-earth MA in $R$Mn$_6$Sn$_6$ and $R$V$_6$Sn$_6$ compounds~\cite{lee2023prb,rosenberg2022prb}, which have recently garnered significant attention as platforms for topological magnets.
For all of these different isostructural series, the calculated easy directions are consistent with experiments, as long as reliable experimental measurements are available for comparison.
Among these two dozen compounds, in addition to the easy-axis and easy-plane anisotropy, some of them exhibit non-trivial easy-cone angles, e.g., $\sim30^\circ$ in Nd$_2$Fe$_{14}$B and $\sim 45^\circ$ in DyMn$_6$Sn$_6$ and HoMn$_6$Sn$_6$.
Moreover, we found that not only the easy directions but also the magnitudes of MA are comparable to existing experiments~\cite{riberolles2022prx,rosenberg2022prb}.
Therefore, our benchmarking of MA in all of these systems validates the usefulness of applying simplistic DFT+$U$ total energy calculations to investigate rare-earth MA, provided that Hund's rules are enforced.

\section{$4f$ anisotropy: Perturbation theory for fast scanning}
Perturbation theory (PT) on top of magnetic force theory has been widely used to calculate and spatially resolve MA in non-$4f$ systems, providing a microscopic understanding of MA.
Since SOC is much smaller than the CF in $d$-electron systems and is treated as a perturbation, one obtains $K=\frac12 K_\text{SO}$ according to second-order perturbation theory.
In other words, the total MA is half of the anisotropy of the SOC energy, $K_\text{SO}$.
Unlike total MA, $K_\text{SO}$ can be resolved into sites, orbitals, spin channels, and bandfillings.

In contrast to $d$-electron systems, in heavy $R$ systems, CF is much smaller than SOC and should be treated as a perturbation.
When the spin rotates, the $4f$ charge is locked to the spin by the dominant SOC and rotates rigidly with the spin.
As a result, the SOC energy $E_\text{SO}$ remains the same during the rotation, and the MA, in principle, can be calculated as $K=K_\text{CF}$ in first-order perturbation theory.

The challenge lies in the accurate estimation of CF energy in open-$4f$-shell elements using DFT+$U$ methods, where CF is overestimated by an order of magnitude, as the aspherical $4f$ charge induces a much larger CF splitting than the ligands.
A quick and rough fix is to use the CF levels of isostructural compounds with $R=$ Gd, whose half-filled $4f$ orbitals give a spherical charge and minimize the CF splitting caused by $4f$ electrons themselves.
Obviously, one should expect that the ligand-only-induced CF splittings would vary across the $R$ series, deviating from the values in the Gd counterpart.
However, even with this rough estimation of CF, we have shown that the perturbation treatment of $4f$ MA provides a good description of MA in $R$Mn$_6$Sn$_6$~\cite{lee2023prb}.
To further demonstrate the validity of PT application on rare-earth anisotropy, we next model the $4f$ MA in $\rco$ and compare it with the $4f$-only contributions obtained from total energy calculations.

\begin{figure}[htb]
\centering
\begin{tabular}{c}
  \includegraphics[width=.6\linewidth,clip]{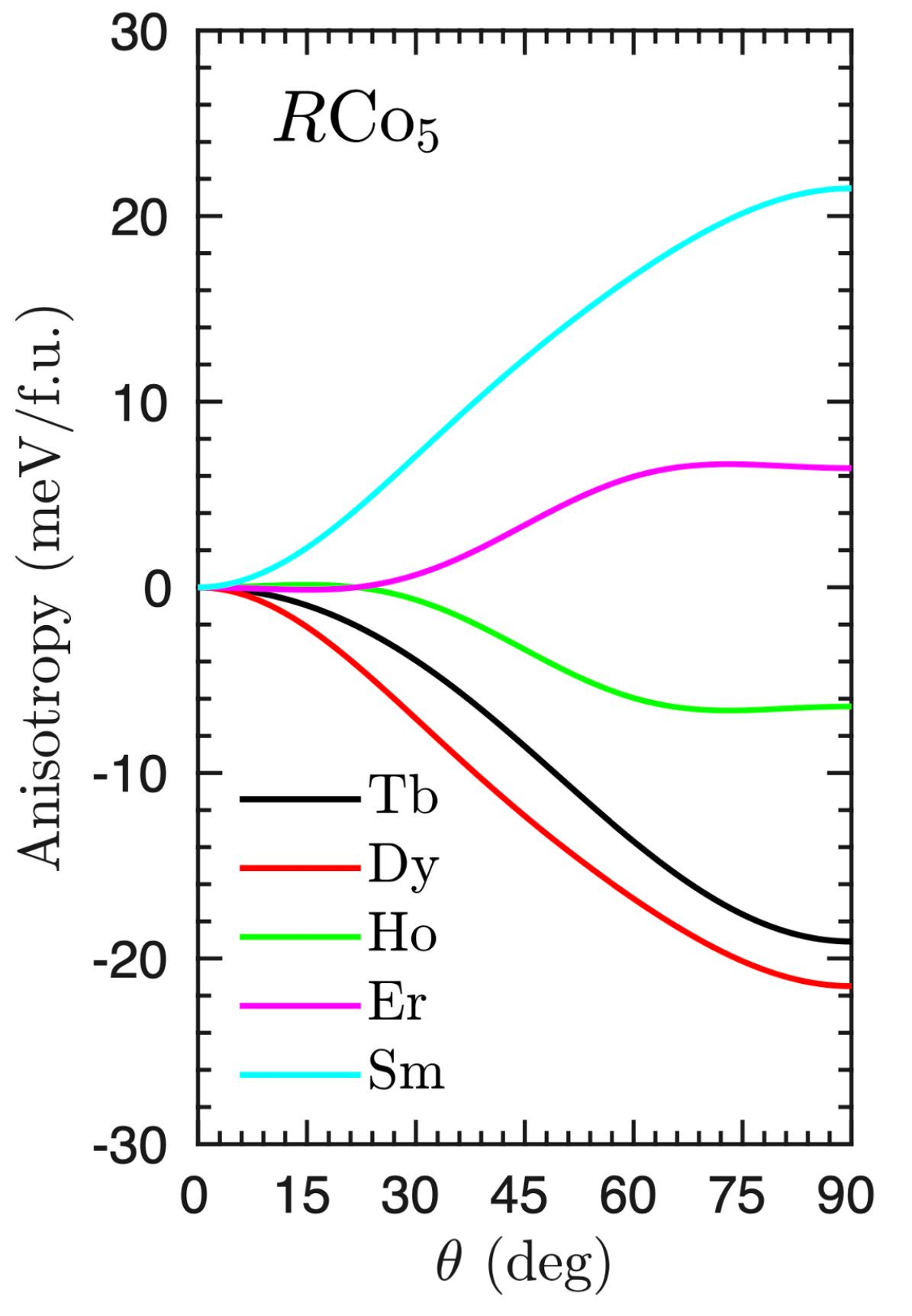}  
\end{tabular}%
\caption{Energy as a function of the spin quantization angle, characterized by the polar angle $\theta$, in $\rco$, modeled using perturbation theory and GdCo$_5$ CF levels. The CF levels of GdCo$_5$ are calculated in DFT+$U$ without SOC and with $U=9.8$ eV.
  Here, we disregard the energy difference between the $\st{m_l=\pm3}$ $4f$ levels, which reflects in-plane MA but not uniaxial MA.
}
\label{fig:ma_model_15}
\end{figure}

\rFig{fig:ma_model_15} shows the $4f$ MA calculated in PT as $E_\text{CF}(\theta,\phi)$ using the GdCo$_5$ CF levels at $\Gamma$ obtained in scalar-relativistic DFT+$U$ calculations:
\begin{equation}
E(\theta,\phi) = \sum_{m\in \text{Occ.}~4f}\mvm{\Psi_m^{\theta,\phi}}{\hCF}{\Psi_m^{\theta,\phi}}. 
\label{eq:ECF}
\end{equation}

Here, $E_\text{CF}(\theta,\phi)$ is obtained by evaluating the original CF Hamiltonian in the rotated wavefunctions, or, equivalently, the rotated CF Hamiltonian by $(-\theta, -\phi)$ in the original wavefunctions.
Due to the high symmetry of the $\rco$ crystal structure, the $\hCF$ is diagonal in the real-spherical-harmonics basis at $\Gamma$.
Therefore, the eigenvalues of the seven occupied $4f$ states at $\Gamma$ in GdCo$_5$ are sufficient to construct the Hamiltonian $\hCF$ in the complex-harmonics basis, which is the natural basis of the SOC Hamiltonian.
The rotated wavefunctions and Hamiltonian can be calculated using the Wigner rotation matrix.
The calculated MA profiles in PT compare well with the total energy calculations shown in \rfig{fig:total_e_ma_15}.
Our results further demonstrate the validity of the PT approach in describing rare-earth MA.

Due to its simplicity, such PT calculations can be used for 1) fast screening of MA and 2) for understanding the origin of rare-earth MA in a system.
For example, large easy-axis MA is required for many applications, such as permanent magnets and topological magnets.
Total energy calculations are more demanding, and special care must be taken to ensure convergence to the desired $4f$ configurations at every spin direction.
In this context, before conducting more reliable total energy calculations, PT calculations can be used for a rapid initial screening of rare-earth MA to identify potential easy-axis rare-earth MA in unexplored crystal structures.
Furthermore, the PT approach can be used to decompose MA contributions into those from different rare-earth sites, such as in $\rfeb$, and analyze how the MA changes with other tuning parameters, thereby aiding in the understanding of the origin of MA in a system.

\section{Conclusions }
In summary, using TbMn$_6$Sn$_6$, we illustrate a general challenge of calculating rare-earth magnetocrystalline anisotropy in DFT and related methods, which often fail to reproduce the correct Hund's rule ground state of rare-earth elements due to significant orbital dependence of the self-interaction error for strongly localized $4f$ orbitals, and the lack of explicit proper orbital polarization treatment.
The true ground state may appear as a metastable state that lies several hundred meV above, resulting in an incorrect $4f$ orbital occupation associated with an incorrect $4f$ charge density, which in turn leads to incorrect magnetocrystalline anisotropy.
However, as the self-interaction error and orbital polarization are, in principle, rotationally invariant, the anisotropy of the true ground state might be expected to remain correct if Hund's rules are enforced by hand.

We have benchmarked this approach on materials with heavy rare-earth atoms with saturated moments where Hund's rules are expected to be satisfied and the single Slater determinant description is suitable.
Notably, in $\rco$, $\rfeb$, and other compounds, the calculated easy directions (including easy axes, planes, and conical angles) have all agreed well with low-temperature measurements.

Besides total energy calculations, we also demonstrate the application of perturbation theory for evaluating rare-earth anisotropy in $\rco$.
The good agreement between the perturbation approach and total energy calculations shows that it can be a useful tool for fast screening of new systems.
Moreover, in analogy to using the SOC anisotropy to spatially resolve $3d$ anisotropy, such perturbation treatment of crystal field energy can be used to resolve anisotropy in systems that contain multiple nonequivalent rare-earth sites, aiding in the understanding the microscopic origin of rare-earth anisotropy.

\acknowledgments

LK gratefully acknowledges discussions with the late Ralph Skomski.
The work at Ames National Laboratory is supported by the U.~S.~Department of Energy (USDOE), Office of Basic Energy Sciences, Division of Materials Sciences and Engineering.
The initial work by LK were supported by the USDOE Early Career Research Program.
Ames National Laboratory is operated for the U.S. Department of Energy by Iowa State University under Contract No. DE-AC02-07CH11358.
IM acknowledges support from DOE under the grant DE-SC0021089.

\bibliography{aaa}

\end{document}